\documentclass[11pt,a4paper]{article}
\usepackage{url}
\usepackage{jcappub}
\usepackage{amsfonts}
\usepackage{subfigure}
\usepackage{graphicx}
\usepackage{appendix}

\newcommand{\be}{\begin{equation}}
\newcommand{\ee}{\end{equation}}
\newcommand{\bea}{\begin{eqnarray}}
\newcommand{\eea}{\end{eqnarray}}

\def\h18{\hbox{H1821$+$643\,}}

\newcommand{\beq}{\begin{equation}}

\newcommand{\eeq}{\end{equation}}
\newcommand{\bef}{\begin{figure}}
\newcommand{\eef}{\end{figure}}

\newcommand{\GeV}{{\rm GeV}}

\begin{document}

\title{On ALP scenarios and GRB 221009A}

\author{Pierluca Carenza,}
\author{M.C.~David Marsh}

\affiliation{The Oskar Klein Centre, Department of Physics, Stockholm University, Stockholm 106 91, Sweden}

\emailAdd{pierluca.carenza@fysik.su.se} \emailAdd{david.marsh@fysik.su.se}

\abstract{
The extraordinarily bright gamma-ray burst GRB 221009A was observed by a large number of observatories, from radio frequencies to gamma-rays. Of particular interest are the reported observations of photon-like air showers of very high energy:  an 18 TeV event in LHAASO and a 251 TeV event at Carpet-2. Gamma rays at these energies are expected to be absorbed by pair-production events on background photons when travelling intergalactic distances. Several works have sought to explain the observations of these events, assuming they originate from GRB 221009A, by invoking axion-like particles (ALPs). We reconsider this scenario and account for astrophysical uncertainties due to poorly known magnetic fields and background photon densities. We find that, robustly, the ALP scenario cannot simultaneously account for an 18  TeV and a 251 TeV photon from GRB 221009A.
}
\maketitle

\section{Introduction}\label{sec:intro}
Recently, a remarkably energetic gamma-ray burst, GRB 221009A, was detected by several telescopes across the electromagnetic spectrum \cite{GCN32632,GCN32636,GCN32637,GCN32642,GCN32648,GCN32658,GCN32677,GCN32686,GCN32757,GCN32765,GCN32793,2022ATel15651....1N,2022ATel15653....1B,2022ATel15655....1F,2022ATel15656....1P,2022ATel15660....1T,2022ATel15662....1P,2022ATel15661....1T,2022ATel15663....1S,2022ATel15664....1I,2022ATel15665....1B,2022ATel15668....1K,2022ATel15669....1D,2022ATel15671....1T,2022ATel15675....1F,2022ATel15677....1K}.  
Optical observations of the afterglow by X-shooter at the Very Large Telescope at the Gran Telescopio de Canarias
determined the redshift of the object to $z=0.151$ \cite{GCN32648, GCN32765, GCN32686}, corresponding  to a luminosity distance of approximately 600 megaparsec (Mpc).  A GRB this bright is rare: the probability of observing a similar event within 50 years has been estimated to approximately $10\%$ \cite{GCN32793}.  

Particularly striking are the reported detections of very high-energy photons by terrestrial observatories. The Large High Altitude Air Shower Observatory (LHAASO) detected the GRB with both the water Cherenkov detector Array (WCDA) and the air shower detector KM2A, and reported more than 5000 associated photons with energies between 500 GeV and 18 TeV arriving within 2000 seconds of the trigger \cite{GCN32677}. A follow-up analysis of data from the Carpet-2 air shower array at the Baksan Neutrino Observatory identified a photon-like air shower event at 251 TeV from the direction of the GRB arriving 4356 seconds after the trigger \cite{Carpet2}. It remains possible that some of the photons detected by LHASSO, and the air-shower reported from Carpet-2, are of galactic origin or are misidentified cosmic rays; however, it is instructive and relevant to explore the implications of these observation assuming the photons originated from GRB 221009A.

A sufficiently energetic photon can annihilate against a low-energy photon to produce an electron-positron pair: this is the Breit-Wheeler process. As a consequence, the mean-free path of high-energy photons 
propagating through a cosmic environment filled with extra-galactic background light (EBL)
decreases rapidly with energy, once this process becomes kinematically accessible
 \cite{Gould:1966pza, Gould:1967zzb, Gould:1967zza}. A simple analytic approximation at low redshift is given by \cite{Salamon:1994un, Mirizzi:2009aj}
 $$
 \Gamma_\gamma^{-1}(E) \approx \frac{1500\, \text{Mpc}}{\kappa} \left(\frac{\text{TeV}}{E} \right)^{1.55} \, ,
 $$
where $1\leq \kappa\leq 2.5$ parameterise the uncertainty in the EBL density. For photons with \mbox{$E\gg $ TeV}, cosmos is opaque. However, tentative hints of excessive long-distance propagation of high-energy photons from active galactic nuclei and blazars have been reported for some time, as reviewed in~\cite{Biteau:2022dtt, Troitsky:2016akf}. If the 251-TeV photon shower detected by Carpet-2 truly emanated from a photon emitted at $z=0.15$, this would provide a clear sign of non-standard physics \cite{Carpet2, Galanti:2022pbg, Troitsky:2022xso}.

An intriguing scenario that makes it easier for very high-energy photons to traverse cosmic distances involves one of the most well-motivated extensions of the Standard Model: axion-like particles (ALPs). ALPs are naturally light pseudo-scalar particles that frequently appear in theories with hidden, broken symmetries or extra dimensions. The phenomenology of ALPs is largely determined by two parameters of the low-energy effective theory: the ALP mass, $m_a$, and the coupling strength to photons, $g_{a\gamma}$, that appear in the Lagrangian 
\begin{equation}
    {\cal L} = - \frac{1}{4} F_{\mu \nu}F^{\mu \nu} - \frac{1}{2} \partial_\mu a \partial^\mu a - \frac{1}{2}m_a^2 a^2 - \frac{g_{a\gamma}}{4} a F_{\mu \nu} \tilde F^{\mu\nu} \, .
    \label{eq:lag}
\end{equation}
Here $\tilde F^{\mu \nu} = \epsilon^{\mu \nu \lambda \rho} F_{\lambda \rho}$. ALPs and photons can interconvert in background magnetic fields, and ALP-photon conversion underpins most experimental proposals to search for ALPs \cite{Irastorza:2018dyq}, as well as many astrophysical searches \cite{ParticleDataGroup:2022pth}. If an ALP exists, a fraction of the electromagnetic flux from a GRB will be converted into ALPs in the magnetic fields of the host galaxy and in the extragalactic magnetic fields (EGMF) along the path of the photon. A subset of these ALPs will travel towards earth and reconvert into photons in the Milky Way (MW) magnetic field. This way, ALPs furnish a new channel for long-distance transmission of high-energy photons that circumvents the Breit-Wheeler process.   

\emph{Does this ALP scenario provide a viable explanation for the observations by LHAASO and Carpet-2?} Previous works on this question include \cite{Galanti:2022pbg,Baktash:2022gnf,Lin:2022ocj,Troitsky:2022xso, Gonzalez:2022opy}.
Reference \cite{Galanti:2022pbg} pointed out that the observation of a 251 TeV photon originating from GRB 2221009A would indicate non-standard physics, and suggested ALPs could explain the event. Reference \cite{Baktash:2022gnf} diligently analysed the reported LHAASO detection and the implications for ALPs. Reference \cite{Troitsky:2022xso} argued that the ALP scenario provide a viable explanation for the reported observations of both an 18 TeV and an 251 TeV photon from GRB 221009A, for a certain range of the ALP parameters. By contrast, ref.~\cite{Gonzalez:2022opy} found, in a limited analysis, that an 18 TeV photon but not the 251 TeV even can be explained by ALPs.  See also \cite{Lin:2022ocj, Nakagawa:2022wwm}.

The ALP predictions depend on poorly constrained astrophysical magnetic fields, and one might fear that this would preclude a definite analysis on the viability of the scenario. In this paper, we model and account for the uncertainties in the relevant astrophysical magnetic fields and the EBL, and find that the ALP predictions are quite robust. The definite answer to the question above is: \emph{no, it is not possible to generate one expected event at 251 TeV at Carpet-2 without simultaneously generating too many photons at LHAASO.} In other words, the ALP explanation of the highest energy events associated with GRB 221009A is not viable.

\subsection{ALP-photon mixing at high energies}
\label{sec:ALPphoton}
In this section, we briefly review the propagation of high-energy photons and ALPs.  
The evolution of the linearised photon-ALP system in a cold, magnetised plasma can be expressed as a Schrödinger-like equation for the state vector $\Psi= (A_x, A_y,a)^{T}$. 
The hermitian Hamiltonian is given by~\cite{Raffelt:1987im}
 \begin{equation}
        H= 
        \begin{pmatrix}
        \Delta_{\gamma_x} & 0 & \Delta_{a \gamma_x} \\ 
        0 & \Delta_{\gamma_y} & \Delta_{a \gamma_y} \\
       \Delta_{a \gamma_x} & \Delta_{a \gamma_y} & \Delta_a    
        \end{pmatrix} \, ,
\label{eq:ham}
    \end{equation}
where
\begin{align}
    \Delta_{a\gamma_i}&= \frac{g_{a\gamma}}{2} B_i \simeq    1.5\times10^{-2} 
\left(\frac{g_{a\gamma}}{10^{-11}\textrm{GeV}^{-1}} \right)
\left(\frac{B_i}{10^{-6}\,\rm G}\right) {\rm kpc}^{-1}
\nonumber\,,\\
    \Delta_a &= -\frac{m_a^2}{2 \omega} \simeq 
 -0.8 \times 10^{-4} \left(\frac{m_a}{10^{-8} 
        {\rm eV}}\right)^2 \left(\frac{\omega}{10^2 \,\ {\rm TeV}} \right)^{-1} 
        {\rm kpc}^{-1}\, ,
\end{align}
where $\omega$ denotes the mode energy.
Due to QED birefringence (the Cotton-Mouton effect), the propagation of the photon modes depend on the photon polarisation, and the matrix elements $\Delta_{\gamma_i}$ can be found by rotation from a local basis in which the magnetic field points along a fixed coordinate axis. The relevant matrix elements are then
\begin{align}
     \Delta_\parallel &=  \Delta_{\rm pl} + \frac{7}{2}\Delta_{\rm QED} +  \Delta_{\rm CMB} \\
     \Delta_\perp &=  \Delta_{\rm pl} + 2 \Delta_{\rm QED} +  \Delta_{\rm CMB}  \, ,
\end{align}
where
\begin{eqnarray}  
\Delta_{\rm pl}&\simeq & 
  -1.1\times10^{-12}\left(\frac{\omega}{10^2 \,\ {\rm TeV}}\right)^{-1}
         \left(\frac{n_e}{10^{-3} \,{\rm cm}^{-3}}\right) 
         {\rm kpc}^{-1}
\nonumber\,,\\
\Delta_{\rm QED}&\simeq & 
6.1\times10^{-4}\left(\frac{\omega}{10^2 \,\ {\rm TeV}}\right)
\left(\frac{B_\text{T}}{10^{-6}\,\rm G}\right)^2 {\rm kpc}^{-1} \nonumber\,, \\
\Delta_{\rm CMB}&\simeq & 
8.0\times 10^{-3}\left(\frac{\omega}{10^2 \,\ {\rm TeV}}\right)
{\rm kpc}^{-1} \,,
\label{eq:Delta0}\end{eqnarray}
where $n_e$ denotes the electron density and $B_\text{T}$ denotes the strength of the magnetic field in the directions transverse to the propagation. The contribution $\Delta_{\rm CMB}$ account for photon-photon dispersion, and the expression for $\Delta_{\rm CMB}$ is a valid approximation for $\omega\lesssim300~\mathrm{TeV}$~\cite{Dobrynina:2014qba, Mastrototaro:2022kpt}. For the high-energy systems we are interested in, $\Delta_{\rm pl}\ll \Delta_{\rm CMB}$. We note that the ratio of $\Delta_{\rm CMB}$ and $\Delta_{\rm QED}$ is independent of energy (in this range), and only depends on the magnetic field; for field strengths ${\cal O}(B_\text{T}) \lesssim \mu$G, the contribution from  $\Delta_{\rm CMB}$ dominates $\Delta_{\rm QED}$: $\Delta_\gamma=\Delta_\parallel\approx \Delta_\perp\approx \Delta_{\rm CMB}$.

When photon absorption is relevant, e.g.~through the Breit-Wheeler process, the system no longer evolves unitarily but is governed by the modified  Schrödinger-like equation
$$
i \frac{d}{dz} \Psi(z) = (H - i D) \Psi(z) \, ,
$$
where 
$$
D(z) =
\begin{pmatrix}
\beta(z) & 0 &0 \\
0 & \beta(z) &0 \\
0 & 0 & 0
\end{pmatrix}
$$
where real $\beta$ encodes the damping. The optical depth for the photons is  given by $\tau(z)/2 = \int_0^z dz'\, \beta(z')$ and the transfer matrix from $z_1$ to $z_2$ is given by
\beq
T_0(z_2, z_1) = e^{-i \int_{z_1}^{z_2} dz'\, \left( H_0 - i D\right)} \, .
\eeq

\begin{figure}[t!]
\centering
\includegraphics[width=0.8\textwidth]{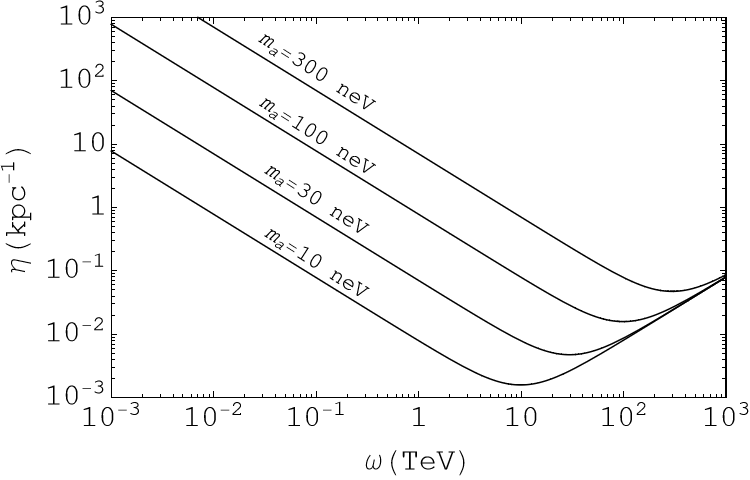}
\caption{Momentum transfer $\eta$ as function of the ALP energy for various ALP masses. As the energy increases, the CMB refraction becomes more important than the ALP mass term and $\eta$ increases linearly with the energy.}
\label{fig:etavsomega}
\end{figure}

When the mixing of ALPs and photons is sufficiently weak, the modified Schrödinger-like equation admits perturbative solutions that can be evaluated  efficiently using Fourier transforms \cite{Raffelt:1987im, Marsh:2021ajy,Carenza:2022zmq}.
The perturbative amplitudes for each linear polarisation, including the effect of damping,  are up to an irrelevant phase given by \cite{Marsh:2021ajy} 
\begin{align}
{\cal A}_{\gamma_i \to a} &= - i \frac{g_{a\gamma}}{2} \int_{-\infty}^{\infty} dz' e^{- i\eta z'}\, e^{-\tau(z')/2} B_i(z') \nonumber \\
{\cal A}_{a \to \gamma_i} &= 
- i \frac{g_{a\gamma}}{2} e^{-\tau_\infty/2} \int_{-\infty}^{\infty} dz' e^{i \eta z'}\, e^{\tau(z')/2} B_i(z')
\, ,
\label{eq:pertA}
\end{align}
where $\eta= \Delta_{\rm CMB}- \Delta_a>0$. In \eqref{eq:pertA}, we take the source to be located at $z'=0$, and the function $B_i(z')$ to denote the relevant magnetic field when $z'>0$ and to vanish when $z'<0$. Here $\tau_\infty$  is the total optical depth relevant for the re-conversion process: in the ALP scenario we consider, it is the optical depth of the MW along a given sightline. 
Clearly, at a fixed energy, the  conversion amplitudes are given by a single Fourier mode, ${\cal A}(\eta)$, of $e^{\mp\tau/2} B_i$. The parameter $\eta$ can be interpreted as the momentum transfer, and is plotted as a function of the mode energy in fig.~\ref{fig:etavsomega}. For energies  $10\, \text{TeV} \lesssim \omega \lesssim 251\, \text{TeV}$ and ALP masses $m_a\lesssim 100\, \text{neV}$, the parameter $\eta^{-1}\gtrsim {\cal O}(10\, \text{kpc})$. This means that, in this parameter range, galactic ALP-photon conversion is sensitive only to the longest-wavelength coherent structure of the magnetic field, and not the detailed substructure. Moreover, the power spectrum of EGMF are expected to peak on scales of ${\cal O}(\text{Mpc})$, but for $m_a \gtrsim 30\, \text{neV}$,  $\eta^{-1}$ saturates at a shorter distance scale, leading to a suppressed conversion probability. Note also that for energies around 251 TeV, the value of $\eta$ becomes mass-independent below $m_a\lesssim 100$ neV. This means, in particular, that the intergalactic conversion probability does not increase as $m_a$ decreases (contrary to the analysis of \cite{Troitsky:2022xso}, in which photon-photon dispersion was omitted).  

By direct comparison to the full numerical solutions in several examples, we have found that the perturbative formalism is applicable for $g_{a\gamma}\lesssim3\times10^{-11}\, \text{GeV}^{-1}$ for the magnetic field models and ALP masses that we will be interested in, and below this limit,  it offers an efficient way to calculate the conversion probability when scanning a large number of possible magnetic field profiles. We note that the Breit-Wheeler mean free path exceeds galactic scales for the relevant energy range, and the factors involving $e^{\pm \tau/2}$ can be dropped from the amplitudes when considering conversion in the MW or the host galaxy. Moreover, the unpolarised conversion probabilities are given by averaging over polarisations: \mbox{$P_{\gamma \to a} = \tfrac1 2 \left(P_{\gamma_x \to a} + P_{\gamma_y \to a} \right)$}.

\begin{figure}[t!]
\centering
\includegraphics[width=0.8\textwidth]{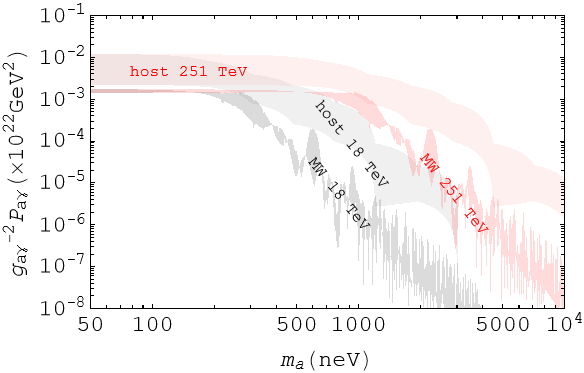}
\caption{Conversion probability as function of the ALP mass for two different energies, 18~TeV (black) and 251~TeV (red), in the MW and in the source host galaxy. The bands are estimated as described in the text. Note that this result is obtained in the perturbative regime and is valid for $g_{a\gamma}\lesssim 3\times10^{-11}\GeV^{-1}$.}

\label{fig:Bfields}
\end{figure}

\subsection{Magnetic field modelling}
\label{sec:B}
%%%%%%%%%%%%%%%%%%%% Start pasted

In order to determine the robustness of the predictions of the ALP scenario, one must investigate the sensitivity of the predictions to the modelling of the astrophysical magnetic fields. Conversion of photons into ALPs can happen in jet region near the source ($z\lesssim 10^{-5}$ pc), the host galaxy ($z\lesssim 20$ kpc), and intergalactic space ($z\gtrsim$ Mpc). Re-conversion of ALPs into photons is assumed to happen in the MW. Out of these, only the MW magnetic field is well-constrained by observations. We now discuss our modelling of each of these magnetic fields in turn.

The mixing of  photons and ALPs in the high-density ($n_e \sim 10^8\, \text{cm}^{-3}$), strongly magnetised ($B \sim 10^8\,$G) jet region has been considered in e.g.~\cite{Mena:2011xj}. At TeV energies however, photon-to-ALP conversion is suppressed due to the very large value of $\Delta_{\rm QED}$ and we will not consider this contribution further. 

Galactic magnetic fields of the order of $\mu$G coherent over kiloparsec scales have been observed in a large number of galaxies. The magnetic field of the host galaxy of GRB 221009A is of key importance to the ALP scenario, but is not directly constrained by observations. The models considered so far in the literature are simplistic: ref.~\cite{Baktash:2022gnf} considered an $0.5\, \mu$G magnetic field coherent over 10 kpc, while ref.~\cite{Troitsky:2022xso} assumed the host galaxy to be sufficiently magnetised to yield maximal mixing between photons and ALPs. 

In this paper, we instead model the host galaxy by using the best-known galactic magnetic field: that of the MW. A large number of studies of Faraday rotation measures and polarised synchrotron emission constrain the MW magnetic field and have led to detailed models. Most notable are the Jansson \& Farrar (JF) model \cite{Jansson:2012pc}, and its updated version that also accounts for data from the Planck satellite \cite{Adam:2016bgn}. The JF model includes a regular and a turbulent component, however, we neglect the latter since it has little impact on conversion rates and does not affect our conclusions~\cite{Carenza:2021alz}. We assume the source to be centrally located within the host galaxy, 
with a magnetic field given by \cite{Adam:2016bgn}, and simulate 1600 possible lines-of-sight for 30 mass-points between $50$ neV and $10^4$ neV using the perturbative formalism described in section \ref{sec:ALPphoton}.\footnote{We have verified that the perturbative results are in excellent agreement with the non-perturbative numerical solution as long as $g_{a\gamma} \lesssim 3\times 10^{-11}\, \text{GeV}^{-1}$.}
Figure \ref{fig:Bfields} shows the 1$\sigma$ bands for the host galaxy conversion probabilities at $\omega=18$ TeV and $\omega= 251$ TeV as a function of the ALP mass. One standard deviation corresponds to a change in the conversion probability by approximately $\pm 67\%$.  We note that the simplistic model of \cite{Baktash:2022gnf} provides a rather typical example of the conversion probabilities for sufficiently small ALP masses, but is less reliable for $m_a\gtrsim 300$ neV.   Maximal mixing (cf.~\cite{Troitsky:2022xso}) does not occur in this region of the parameter space. 

ALP-photon mixing may also occur in the intergalactic space between the host galaxy and earth. There are indirect arguments for a lower bound on EGMF in voids based on secondary inverse Compton photons emitted by Breit-Wigner electrons and positions created from TeV gamma rays from Blazars: this gives $B \gtrsim 10^{-16} \, \text{G}$ on megaparsec scales \cite{Podlesnyi:2022ydu}. An upper bound on the EGMF can be derived from bounds on the diffuse radio background, giving $B \lesssim {\cal O}(\text{nG})$. In this work, we will consider both the case of negligible extragalactic conversion, and then a the case of a close-to-maximal EGMF of nano-Gauss strength over megaparsec distances, as in \cite{Galanti:2022pbg}. Following ref.~\cite{Mirizzi:2009aj}, we model this EGMF as a cell model, with a coherence length of 1~Mpc and an average magnitude of the magnetic field equal to 1~nG. For this model, the average conversion probability is obtained analytically~\cite{Mirizzi:2009aj}.

Finally, the MW  magnetic field supports ALP-to-photon re-conversion. To estimate the related uncertainty of the ALP scenario we fix the sight-line towards GRB 221009A and  
consider the original JF model with parameters given in ref.~\cite{Jansson:2012pc}, giving a lower bound for ALP-photon conversion probabilities, and the JF model with the updated parameters given in Table~C.2 of~\cite{Adam:2016bgn} (i.e.~``Jansson12c'', ordered fields), giving an upper bound. 
We estimate the uncertainty in the MW prediction of the re-conversion probability to $\pm10\%$.
Figure~\ref{fig:Bfields} shows the MW conversion probabilities at $\omega= 18\, \text{TeV}$ and $\omega= 251\, \text{TeV}$ as function of the ALP mass. 
 Note that our assumption that GRB 221009A was centrally located within the host galaxy leads to larger conversion probabilities in the host galaxy than in the MW. For a peripherally located GRB, the conversion probabilities extend to significantly lower values, and it would be even more difficult to realise the ALP scenario for GRB 221009A. 

The ALP parameter space of interest for explaining GRB 221009A extends beyond the region where perturbation theory is applicable, but determining the uncertainty in the ALP predictions non-perturbatively is not efficient. We therefore use the perturbative results to model the uncertainty across the parameter space. We solve for a single line-of-sight and use multiple magnetic field strengths to estimate the non-perturbative host galaxy uncertainty. The $67\%$ uncertainty in the host galaxy prediction for the conversion probability translates into a $\sim40\%$ uncertainty in the magnetic field in the perturbative limit. We vary the magnetic field along the sightline by $\pm50\%$ as an estimate of the uncertainty.

%%%%%%%%%%%%%%%%%%%%%%%%%%%%%%%%%%%%%%%%%%%%%%

\begin{figure}[t!]
\centering
\includegraphics[width=0.7\textwidth]{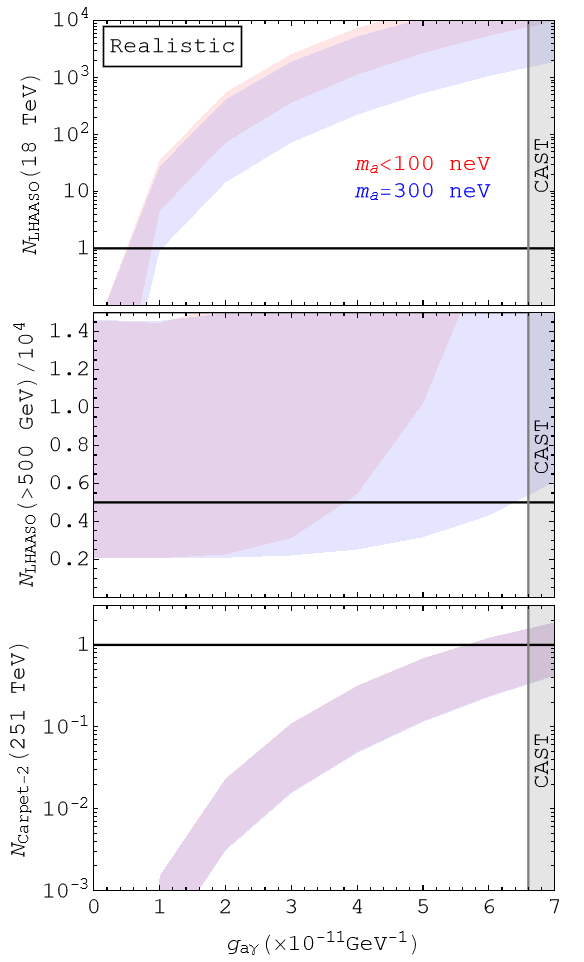}
\caption{Expected number of events 
as function of the ALP-photon coupling
for light ALPs with $m_{a}\lesssim100$~neV (red), and $m_{a}=300$~neV (blue) in the `realistic scenario'. The figures show LHAASO at $18\, (\pm3.6)$~TeV (top); LHAASO from 500~GeV to 21 TeV (middle); and Carpet-2 at $251\, (\pm 50)$~TeV (bottom). Reported observations are marked by black horizontal lines. The gray region indicates the upper limit on $g_{a\gamma}$ set by CAST~\cite{CAST:2017uph} (the same region is also excluded by the Horizontal Branch constraint~\cite{Ayala:2014pea}).}
\label{fig:events1}
\vspace*{1 cm}
\end{figure}
\pagebreak

\begin{figure}[t!]
\centering
\includegraphics[width=0.7\textwidth]{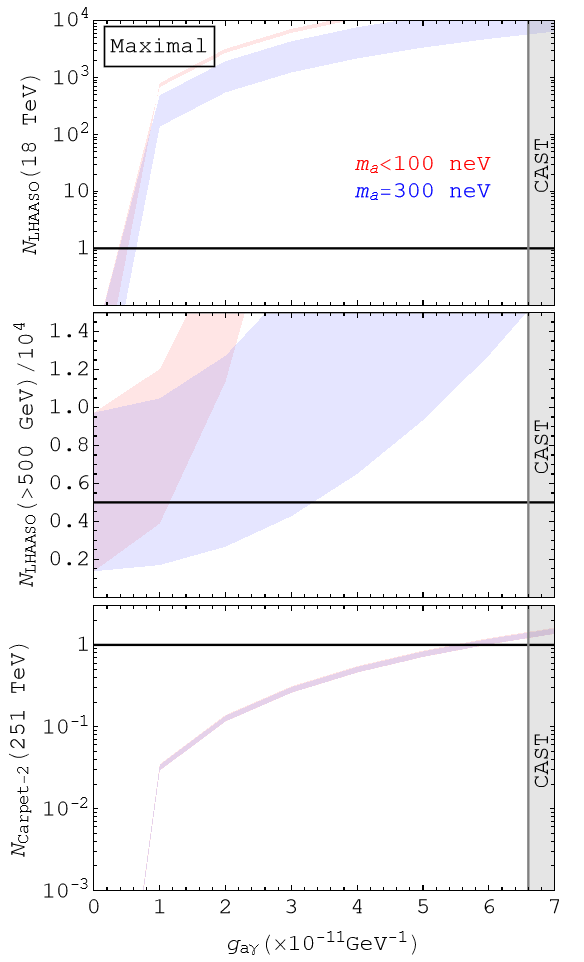}
\caption{Expected number of events 
as function of the ALP-photon coupling in the `maximal' scenario. See fig.~\ref{fig:events1} for explanations. }
\label{fig:events2}
\vspace*{2cm}
\end{figure}
%\pagebreak

\begin{figure}[t!]
\centering
\includegraphics[width=0.7\textwidth]{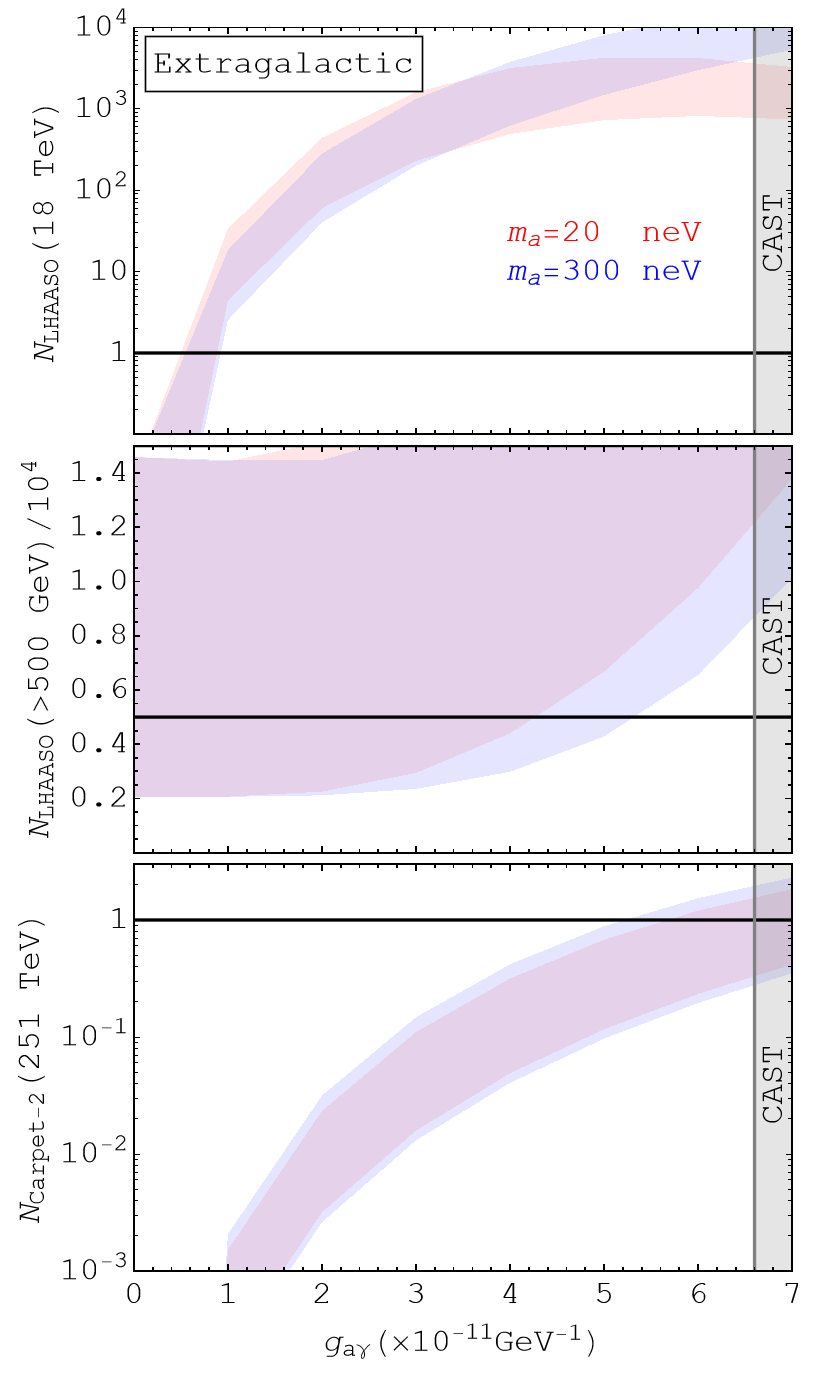}
\caption{Expected number of events 
as function of the ALP-photon coupling in the `extragalactic' scenario. See fig.~\ref{fig:events1} for explanations. The low ALP mass case here refers to $m_{a}=20$~neV.}
\label{fig:events3}
\vspace*{2.5cm}
\end{figure}

\begin{figure}[t!]
\centering
\includegraphics[width=0.7\textwidth]{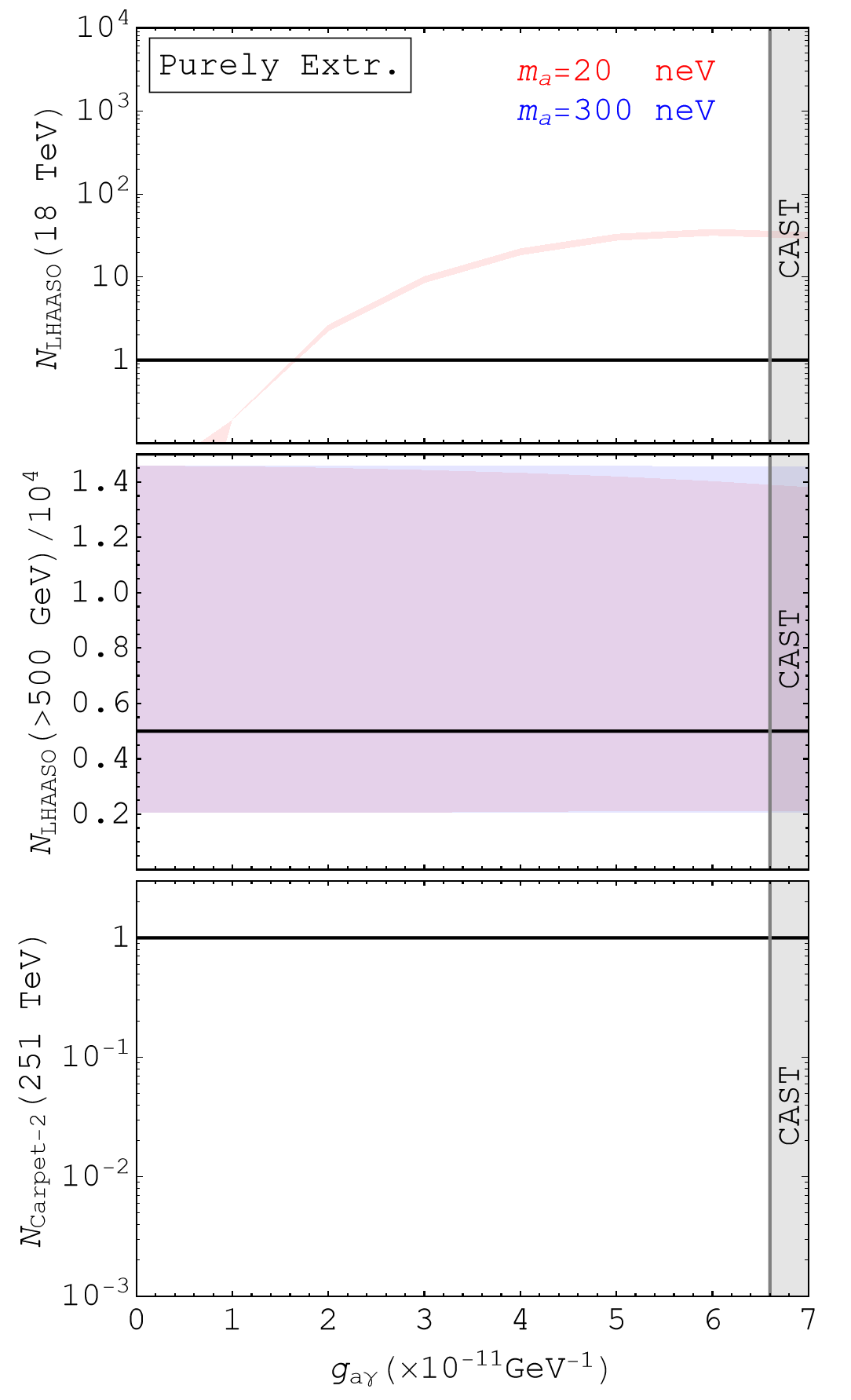}
\caption{Expected number of events 
as function of the ALP-photon coupling in the `purely extragalactic' scenario. See fig.~\ref{fig:events1} for explanations. The low ALP mass case here refers to $m_{a}=20$~neV.}
\label{fig:events4}
\vspace*{2.5cm}
\end{figure}
%\pagebreak

\subsection{Is the ALP scenario viable?}
\label{sec:results}
In this section, we re-examine the ALP-photon predictions for GRB 221009A and determine if there are any ALP parameters that can explain all observations.

In \cite{GCN32658}, Fermi-LAT reported on a preliminary analysis of the photon spectrum in the energy range $100 \, \text{MeV}<\omega< 1\, \text{GeV}$, collected between 200 and 800 seconds after the trigger: 
\begin{equation}
    F(E)=2.1\times10^{-6}\, \left(\frac{E}{1~{\rm TeV}}\right)^{-1.87}\, {\rm cm}^{-2}{\rm s}^{-1}{\rm TeV}^{-1}\,.
    \label{eq:Fermispectrum}
\end{equation}
Following \cite{Baktash:2022gnf}, we regard this spectrum as intrinsic and extrapolate it to the entire range of interest. We note that possible spectral breaks or the decrease of the luminosity with time would reduce the number of very high-energy photons, and make the ALP scenario even more difficult to realise. Thus, for our purposes,  the spectrum \eqref{eq:Fermispectrum} is conservative. 

We model the optical depth following ref.~\cite{Cirelli:2010xx}. This model accounts for various contributions to the low-energy photon background. 
For energies above $\sim300$~TeV, the well-studied Cosmic Microwave Background (CMB) is the main target for the Breit-Wheeler process.
For energies  $10\, \text{TeV} \lesssim \omega \lesssim 300$~TeV, the principal source of absorption is the secondary infrared radiation from dust scatterings. At lower energies, the absorption is dominated by the ultraviolet (UV) light emitted by stars in the low-redshift Universe. The UV component is not well-constrained observationally and, following ref.~\cite{Cirelli:2010xx}, we use two models for the UV component based on ref.~\cite{Dominguez:2010bv} to estimate the uncertainty.  The first model, which we denote `min UV', reproduces the UV photon densities inferred from blazar studies~\cite{MAGIC:2008sib}. The second model, `max UV', rescales the UV photon densities of the minimal model by a factor $1.5$, and  is expected to give an upper limit on possible systematic errors. Clearly, `max UV' predicts more absorption for gamma-rays than `min UV'.

We model the detectors by using the effective areas given in ref.~\cite{Ma:2022aau} for LHAASO and in ref.~\cite{Dzhappuev:2020bkh} for Carpet-2. The exposure time of LHAASO is taken to be 2000~s~\cite{GCN32677} and 4536~s for Carpet-2~\cite{Carpet2,Troitsky:2022xso,CarpetATel-GRB}. We are interested in three observationally constrained quantities:
\begin{itemize}
    \item \# events in LHAASO with $0.5 \, \text{TeV} \leq \omega\leq 21\, $TeV ~~[observation: $\gtrsim 5000$]
    \item \# events in LHAASO with $\omega\approx 18\, (\pm3.6) $TeV ~~~[observation: 1]
    \item \# events in LHAASO with $\omega\approx 251\, (\pm 50)\, $TeV ~~[observation: 1]
\end{itemize}
Here, we have assumed an energy resolution of $\sim 40\%$ for both LHAASO \cite{Baktash:2022gnf, Ma:2022aau} and Carpet-2.

Without invoking non-standard physics, but accounting for the uncertainties in the optical depth, our model predicts between $2\times10^{3}$ and $1.4\times10^{4}$ events above 500~GeV in LHAASO and none of these at around 18~TeV,\footnote{It should be noted though that due to the large uncertainties in the EBL models, the probability of observing an 18 TeV photon at LHAASO is not necessarily very small \cite{Baktash:2022gnf,Zhao:2022wjg}. } as well as no events in Carpet-2. We would now like to determine if any ALP scenario can explain these observations by predicting an expected number of events compatible with the observations. We consider four scenarios for the relevant magnetic fields. \vspace{4 pt}

{\bf The realistic scenario:} photon-to-ALP conversion happens in the galactic magnetic field of the host galaxy of GRB 221009A, which is modelled as discussed in section \ref{sec:B}.  \vspace{4 pt}

{\bf The maximal scenario:} following ref.~\cite{Troitsky:2022xso}, photon-to-ALP conversion happens in the galactic magnetic field of the host galaxy of GRB 221009A, which is assumed to be strong enough that the conversion probability takes the maximal value of $1/3$ at any energy, mass and coupling~\cite{Csaki:2001yk}.  \vspace{4 pt}

{\bf The extragalactic scenario:} photon-to-ALP conversion happens both in the host galaxy, which is considered to be the same of the `realistic' scenario, and in the extragalactic magnetic field, which is assumed to be close to the upper observational bound, as discussed in section \ref{sec:B}. 

{\bf The purely extragalactic scenario:}  photon-to-ALP conversion is assumed to happen only in the extragalactic magnetic field, which is modelled as in the `extragalactic' scenario. This is similar to version 1 of ref.~\cite{Galanti:2022pbg}, which however also stated that the near-source contribution would be appreciable, contrary to our discussion. \vspace{4 pt}

In all scenarios, ALP-to-photon re-conversion in assumed to occur in the MW.

Our results are summarised in figs.~\ref{fig:events1}--\ref{fig:events4} for the four different scenarios. The plots show the predicted number of events in LHAASO (around 18 TeV and above 500 GeV, respectively) and Carpet-2 (around 251 TeV) as a function of the ALP-photon coupling, $g_{a\gamma}$. Since the predictions become independent of the ALP mass for sufficiently small ALP masses, we show only the representative cases of  $m_a<100\, \text{neV}$ and $m_a=300\, \text{neV}$ for the first two scenarios. In the cases involving conversions in the intergalactic medium we consider $m_{a}=20$~neV as representative case for light ALPs.
The conversion probability, and therefore the expected number of events, further decreases with increasing mass, as indicated in fig.~\ref{fig:Bfields}.

In the realistic scenario of fig.~\ref{fig:events1}, the event at 18~TeV detected by LHAASO can be explained by introducing ALPs with $g_{a\gamma}\sim(0.5$--$1)\times10^{-11}\GeV^{-1}$. At this coupling, the prediction for the total number of LHAASO events is essentially unchanged from the no-ALP case. Explaining the Carpet-2 event at higher energy requires a much larger coupling, $g_{a\gamma}\gtrsim5.6\times10^{-11}\GeV^{-1}$, which, however, results in a significantly larger number of expected events in LHAASO  at 18 TeV (and for low masses, also  above $500$ GeV). Thus, the realistic ALP scenario is not a viable explanation for both the LHAASO and Carpet-2 observations.

The maximal scenario of fig.~\ref{fig:events2} is qualitatively similar to the realistic scenario, but with a more narrow confidence band since there is, by assumption, no uncertainty associated with the host galaxy, and the photon flux is reduced by one-third. The couplings required to explain the 18 TeV event at LHAASO is again discrepant with the value required to explain the 251 TeV event at Carpet-2, and the ALP scenario is not viable.

In the extragalactic scenario (fig.~\ref{fig:events3}), the ALP survival probability decreases with energy in the energy range from $500$~GeV to $18$~TeV for light ALPs ($m_{a}\lesssim 20$~neV). For heavier ALPs, the conversion probability is suppressed. As commented in ref.~\cite{Galanti:2022xok}, at $251$~TeV the ALP conversion in the intergalactic medium is inefficient because of the refraction on CMB photons. The prediction for the number of events in Carpet-2 reduces to the `realistic' case.

Finally, in the purely extragalactic scenario in fig.~\ref{fig:events4}, the events detected by LHAASO above $500$~GeV roughly coincides with the  no-ALP case since the extragalactic conversion is quite inefficient. For sufficiently light ALPs and large couplings, a small fraction of photons convert into ALPs, slightly reducing the total number of events detected by LHAASO. For light ALPs ($m_{a}\lesssim 20$~neV) and  $g_{a\gamma}\sim 1.6\times10^{-11}\GeV^{-1}$, the conversion probability becomes high enough to explain the $18$~TeV LHAASO event. Still, the photon-ALP conversion at $251$~TeV is highly suppressed, making impossible for photons to be detected in Carpet-2.

Clearly, the emerging picture is that, even when accounting for the large astrophysical uncertainties, a common ALP explanation of the observations at LHAASO and Carpet-2 of GRB 221009A is strongly disfavored. On the one hand, an ALP introduced to justify the $18$~TeV event in LHAASO would not lead to a significant number of events at much higher energies. On the other hand, an  ALP introduced to justify the Carpet-2 observation is incompatible with the LHAASO observation since it would overproduce gamma-ray events above $500$~GeV and, especially, at $18$~TeV.

\subsection{Conclusions}
We have shown that the previously proposed ALP scenario for explaining the tentative measurements by LHAASO and Carpet-2 of GRB 221009A is not viable. We have accounted for astrophysical uncertainties in the ALP predictions and in the optical depth, but found that ALP scenarios do not simultaneously predict one 18 TeV photon event at LHAASO and a 251 TeV photon at Carpet-2 .  Consistently with \cite{Baktash:2022gnf}, we find that the LHAASO observations can be explained by (but do not unambiguously call for \cite{Baktash:2022gnf,Zhao:2022wjg}) an ALP with a coupling of $g_{a\gamma}\sim(0.5$--$1.6)\times10^{-11}\GeV^{-1}$ for a broad range of masses. Disregarding the reported Carpet-2 event (which may be due to galactic sources or a misidentified cosmic ray), such an ALP is in tension with magnetic White Dwarf polarization measurements~\cite{Dessert:2022yqq}, and at very low masses, with X-ray constraints on ALPs \cite{Marsh:2017yvc, Reynolds:2019uqt, Reynes:2021bpe}.

\subsection*{Acknowledgements} 
The work of PC and DM is supported by the European Research Council under Grant No.~742104 and by the Swedish Research Council (VR) under grants  2018-03641 and 2019-02337.

\bibliographystyle{bibi}
\bibliography{sample.bib}

\end{document}